\documentclass[aps,twocolumn,superscriptaddress,reprint]{revtex4-1}
\usepackage{graphicx}
\usepackage{setspace}
\usepackage{amsmath}
\usepackage{amssymb}
\usepackage{color}
\usepackage{array}
\usepackage{subfigure}
\usepackage{hyperref}
\usepackage{float}
\usepackage{lipsum}
\usepackage{ulem}

\usepackage[all]{xy}
\newcommand{\RN}[1]{%
  \textup{\uppercase\expandafter{\romannumeral#1}}%
}

\begin{document}
%\begin{CJK*}{GB}{}
%\title{Real time dynamics of $1+1$-dimension fermionic quantum field theory}
\title{Simulation of state evolutions in Gross-Neveu model by matrix product state representation}

\author{De-Sheng Li}
%\email[email: ]{lideshengjy@126.com}
\affiliation{College of Physics, Mechanical and Electrical Engineering, Jishou University, Jishou 416000, P.R.China}
\affiliation{Interdisciplinary Center for Quantum Information, National University of Defense Technology, Changsha 410073, P.R.China}

\author{Hao Wang}
%\email[email: ]{}
\affiliation{Department of Physics, National University of Defense Technology, Changsha 410073, P.R.China}

\author{Chu Guo}
%\email[email: ]{}
\affiliation{Quantum Intelligence Lab, Supremacy Future Technologies, Guangzhou 511340, P.R.China}

\author{Ming Zhong}
\email[email: ]{zhongm@nudt.edu.cn}
\affiliation{Department of Physics, National University of Defense Technology, Changsha 410073, P.R.China}

\author{Ping-Xing Chen}
%\email[email: ]{pxchen@nudt.edu.cn}
\affiliation{Interdisciplinary Center for Quantum Information, National University of Defense Technology, Changsha 410073, P.R.China}
\affiliation{Department of Physics, National University of Defense Technology, Changsha 410073, P.R.China}

\begin{abstract}
A quantum algorithm to simulate the real time dynamics of two-flavor massive Gross-Neveu model is presented in Schr$\ddot{o}$dinger picture. We implement the simulation on a classic computer by applying the matrix product state representation. The real time evolutions of up to four particles on a site in initial state are figured out in space-time coordinate. The state evolutions are effectively affected by fermion mass and coupling constant of the model. Especially when the mass of fermion is small enough and the coupling is strong enough, the fundamental fermions evolve synchronistically in space from the two-fermion and four-fermion initial states. These are also the conditions on which the bound states made up of fundamental fermion pairs were found to arise automatically in the literatures. 

\end{abstract}
\maketitle
%\end{CJK*}

\section{Introduction}
\label{intro}
In quantum field theories (QFT), the non-linear interactions among fields usually lead to intricate calculations. Especially when the interaction becomes strong, the perturbative computation is incapable and non-perturbative methods have to be resorted to. A promising attempt at non-perturbative calculation of QFT is quantum computation whose complexity is polynomial rather than exponential as compared to the classic computing. 

There have been concrete progresses to this way. Some quantum algorithms \cite{Tim2006Simulating, Jordan2011Quantum,  Jordan2014Quantum, wiese2014towards, li2018digital, raychowdhury2018tailoring, lamm2019quantum} and quantum simulation methods \cite{Casanova2011Quantum, longhi2012realization, zohar2013cold, tagliacozzo2013simulation,hauke2013quantum, rico2014tensor, mezzacapo2015non, yang2016analog, zohar2017digital, bender2018digital, klar2019quantum} were worked out. After discretization, a QFT will be transferred to a quantum many-body theory. Studies on the quantum simulation of many-body problems \cite{zhang2017observation, bernien2017probing, smith2019simulating} enforce the prospects for simulating QFT in quantum computers. There have also been some reports on experimental realizations, in trapped ion or superconductivity quantum computers, of quantum simulation on some simple QFTs, such as Schwinger model, Yukawa coupling and gauge field theory \cite{Martinez2016real-time, Xiang2018Experimental, klco20192}. 

The interested quantities in usual application of QFT are the S-matrix elements constructed by explicit initial and final particles which are treated as asymptotic states while the real-time evolutions of particles are usually ignored. To have a more complete understanding of QFT, it is worthwhile figuring out its real time dynamic. This can afford us some essential knowledge on the states evolving from the initial time by the interactions, and especially illustrations on how the fundamental particles clump together to form a bound state. This may help us understand the natures of the strong interactions.

In this work, we report a quantum algorithm developed in Schr$\ddot{o}$dinger picture and space-time coordinate to simulate the real time dynamics of two-flavor massive Gross-Neveu model. We implement the algorithm on a classical computer by using the computing routine of matrix product state (MPS). The real time evolutions from up to four particles on a site in initial state are figured out in space-time coordinate. It is observed that the fundamental fermions evolve synchronistically in space from the two-fermion and four-fermion initial states when the mass of fermion is small enough and the coupling is strong enough. These are also the conditions on which the bound states made up of fundamental fermion pairs were found to arise automatically in the seminal Nambu$-$Jona-Lasinio model \cite{Nambu1961tp, Nambu1961fr} and its $(1+1)$-dimentional version, the Gross-Neveu model \cite{gross1974dynamical}. Though the particle states of this work are different from that of \cite{gross1974dynamical} in definition, this simulation might give a series of images on bound state formation. It may provide us an intuitive way to identify the bound states in view of fundalmental particles and the distribution function of the fundamental particles can be extracted from the evolution.

We introduce the quantum algorithm of Gross-Neveu model in Sec.~\ref{alg}. The Gross-Neveu model is a renormalizable $(1+1)$-dimension quartic fermion interaction theory. It was revealed to has dynamical symmetry breaking by a large-N calculation \cite{gross1974dynamical} and therefore is a toy model for quantum chromadynamics and superconductivity. Because of this, it has drawn much attention in high energy and condensed matter physics. Very recently, a quantum algorithm was developed to calculate the correlation functions of the model in \cite{moosavian2018faster}. The MPS calculation results of the algorithm are discussed in Sec.~\ref{result}. As a computation tool developed to run quantum simulations through classical data compression of the quantum states \cite{Schollwock2011}, the MPS has been employed to simulate the dynamics of lattice field theories, special infinite chains, quantum phase transition and correlation functions etc \cite{zohar2017digital,banuls2013matrix,  banuls2017density, buyens2014matrix, banuls2009matrix, wolf2006quantum, cirac2009renormalization, moosavian2019site} and has become a standard numerical tool to simulate one-dimensional quantum many-body systems for ground state searches and time evolution tasks\cite{Vidal2003, DaleyVidal2004, WhiteFeiguin2004, wall2012out-of-equilibrium, jaschke2017open}. We adopt the MPS method to realize running the algorithm in a classic workstation. All results of state evolutions are shown in a series of figures with 11 sites and 150 time slices. A summary is given at the end.

\section{Quantum algorithm of Gross-Neveu model}
\label{alg}
The Lagrangian density of two-flavor massive Gross-Neveu model is written as
\begin{eqnarray}\label{L01}
\mathcal{L}_{GN}=\bar{\psi}_{i}(i\gamma^{0}\partial_{t}+i\gamma^{1}\partial_{x}-m)\psi_{i}+\frac{1}{2}g^{2}(\bar{\psi}_{i}\psi_{i})^{2},
\end{eqnarray}
with the flavor summation index $i$, mass $m$ and coupling constant $g^2$. Obviously it has a global $SU(2)$ flavor symmetry. The fermion mass is present since it can be dynamically produced by spontaneous symmetry breaking in the massless model \cite{gross1974dynamical}. 

In Schr$\ddot{o}$dinger picture, we decompose $\psi_{i}(x)$ in position space \cite{kuypers,nason1984lattice}
\begin{eqnarray}\label{psi}
\psi_{i}(x)=\left(\begin{array}{c}  \hat{a}_{x,i}\\   \hat{b}_{x,i}\end{array}\right),
\end{eqnarray}
where $\hat{a}_{x,i}$ and $\hat{b}_{x,i}$ are annihilation operators of spin up and spin down flavor-i fermions located in position $x$. The particle states of the theory are defined by the creation operators and annihilation operators
\begin{eqnarray} \label{bdagger0}
\hat{a}^{\dagger}_{x,i}|0\rangle_{x,1,i}=|1\rangle_{x,1,i},\quad && \hat{a}^{\dagger}_{x,i}|1\rangle_{x,1,i}=0,\nonumber\\
\hat{a}_{x,i}|1\rangle_{x,1,i}=|0\rangle_{x,1,i},\quad && \hat{a}_{x,i}|0\rangle_{x,1,i}=0,\nonumber\\
\hat{b}^{\dagger}_{x,i}|0\rangle_{x,2,i}=|1\rangle_{x,2,i},\quad && \hat{b}^{\dagger}_{x,i}|1\rangle_{x,2,i}=0,\nonumber\\
\hat{b}_{x,i}|1\rangle_{x,2,i}=|0\rangle_{x,2,i},\quad && \hat{b}_{x,i}|0\rangle_{x,2,i}=0,
\end{eqnarray}
where the subscripts $1$ and $2$ denote spin up and down respectively. The gamma matrices are taken in the representation
\begin{eqnarray}
\gamma^{0}=\left(\begin{array}{cc}1 & 0  \\ 0 & -1\end{array}\right),\gamma^{1}=\left(\begin{array}{cc}0 & i  \\ i & 0 \end{array}\right).
\end{eqnarray}
The exact Hamiltonian is
\begin{eqnarray}
H_{GN}=\int dx \left(-i\bar{\psi}_{i}\gamma^{1}\partial_{x}\psi_{i}+m\bar{\psi}_{i}\psi_{i}-\frac{1}{2}g^{2}(\bar{\psi}_{i}\psi_{i})^{2}\right).
\end{eqnarray}

To perform digital quantum simulation, the Gross-Neveu model has to be discretized
\begin{eqnarray}
 \Omega=  \kappa\mathbb{Z}_{\hat{X}},
\end{eqnarray}
where $\Omega$ is 1-dimension discreted lattice points with spacing $\kappa$ and $\hat{X}=Int(X/\kappa)$. The Hamiltonian is rewritten in a discreted form
\begin{eqnarray}\label{H} 
H_{GN}&=&\sum_{x\in  \Omega,i=1,2} \kappa [-i\bar{\psi}_{i}(x)\gamma^{1}\frac{\psi_{i}(x+\kappa)-\psi_{i}(x-\kappa)}{2\kappa} \nonumber\\
&&+m \bar{\psi}_{i}(x) \psi_{i}(x)-\frac{g^{2}}{2} \left(\bar{\psi}_{i}(x)\psi_{i}(x)\right)^{2}].
\end{eqnarray}
There appears fermion doubling and a Wilson term needs to be added by hand to cancel the effects of the redundant fermions
\begin{eqnarray*}\label{HW}
H_{W}=\sum_{x\in  \Omega,i=1,2} \kappa \left[-\frac{r}{2\kappa}\bar{\psi}_{i}(x)\left(\psi_{i}(x+\kappa)-2\psi_{i}(x)\right.\right.\\
\left.\left.+\psi_{i}(x-\kappa) \right) \right],
\end{eqnarray*}
where $0<r\leqslant 1$ is Wilson parameter. By using \eqref{psi}, the total Hamiltonian is
\begin{eqnarray}\label{hamiltonian}\nonumber
H&=&H_{GN}+H_{W} \\  \nonumber 
&=&\sum_{x\in \Omega, i=1,2}\kappa\left[\frac{1}{2\kappa}\left(\hat{a}^{\dagger}_{x,i}\hat{b}_{x+\kappa,i}+\hat{b}^{\dagger}_{x+\kappa,i}\hat{a}_{x,i} -\hat{b}^{\dagger}_{x,i}\hat{a}_{x+\kappa,i} \right.\right.\\  \nonumber 
&&\left.-\hat{a}^{\dagger}_{x+\kappa,i}\hat{b}_{x,i}\right)  \\  \nonumber
&&+\frac{r}{2\kappa} \left(-\hat{a}^{\dagger}_{x,i}  \hat{a}_{x+\kappa,i}-\hat{a}^{\dagger}_{x+\kappa,i}\hat{a}_{x,i} +\hat{b}^{\dagger}_{x,i}  \hat{b}_{x+\kappa,i}+\hat{b}^{\dagger}_{x+\kappa,i}\hat{b}_{x,i}  \right)\\  \nonumber 
&& +(m+\frac{r}{\kappa})\left( \hat{n}_{x,up,i}- \hat{n}_{x,down,i}\right)\\   \label{Htotal}
&&\left.-\frac{g^{2}}{2} (\hat{n}_{x,up,i}- \hat{n}_{x,down,i}
)^{2}\right],
\end{eqnarray}
where
\begin{eqnarray*}
\hat{n}_{x,up,i}=\hat{a}^{\dagger}_{x,i}\hat{a}_{x,i},\quad \hat{n}_{x,down,i}=\hat{b}^{\dagger}_{x,i}\hat{b}_{x,i}
\end{eqnarray*}
are number operators of spin up and spin down fermions with flavor-i. 

The Hamiltonian (\ref{Htotal}) is obviously hermitian. The way that we define the particle states in (\ref{psi}) and (\ref{bdagger0}) makes it exhibit some neat and simple properties which will manifest in the subsequent simulation of particle evolutions: (i) The flavor is conserved. The flavor does not exist in the initial state will not appear in the evolution. The particles of different flavors with the same spin evolve in the same way. (ii) The spin is not conserved. The first term on the right-hand side dictates a particle with a certain spin on a site transfering to another particle with a flipped spin on the adjacent site. The Hamiltonian flips with the spins flip $H\stackrel{a\leftrightarrow b}{\longrightarrow}-H$. (iii) Particles can propagate from one site to another. This can be seen from the first and second terms which account for the transition between two adjacent particles. These terms are kinematic and act to drive the particles to move and evolve. (iv) The last two terms are eigen-operators of the particle states defined in (\ref{bdagger0}). They tend to make the particles stationary and evolutionless. But these two terms have different signs, the effect of keeping particles stable can be offset to extents depending on the values of the parameters $m$ and $g$. 

We now proceed to digitize the Hamiltonian (\ref{Htotal}) by defining one to one mapping between fermion occupation number states and qubits
\begin{equation}
\label{statemapping}
|0\rangle \leftrightarrow |\downarrow\rangle, \quad |{1}\rangle \leftrightarrow |\uparrow\rangle,
\end{equation}
with the two qubits $|\downarrow\rangle=(0,1)^T$ and $|\uparrow\rangle=(1,0)^T$ corresponding to vacuum and one-fermion states respectively in Fock space. The qubits can then present a representation for operators. Using the Jordan-Wigner mapping, we have the representations for the creation and annihilation operators
\begin{eqnarray} \label{bdagger}
\hat{a}^{\dagger}_{x,i}=(\prod_{\alpha} -\sigma^{\alpha}_{z})\sigma^{x,1,i}_{+},&&    \hat{a}_{x,i}=(\prod_{\alpha} -\sigma^{\alpha}_{z})\sigma^{x,1,i}_{-},\\  \label{b3}
\hat{b}^{\dagger}_{x,i}=(\prod_{\alpha} -\sigma^{\alpha}_{z})\sigma^{x,2,i}_{+},&&    \hat{b}_{x,i}=(\prod_{\alpha} -\sigma^{\alpha}_{z})\sigma^{x,2,i}_{-},\\  \label{c3}
\hat{n}_{x,s,i}=\frac{1}{2}(\sigma_{z}^{x,s,i}+I),
\end{eqnarray}
with $\alpha\in \{ (x^{\prime},s^{\prime},i^{\prime})| \mathcal{K}(x^{\prime},s^{\prime},i^{\prime})<\mathcal{K}(x,s,i)\}$, $s\in \{1,2\}=\{up,down\}$ and the primary key $\mathcal{K}(x,s,i)\in\mathbb{N}^{*}$. The Pauli matrices $\sigma^{x,s,i}$ is the operation acting on $\mathcal{K}(x,s,i)$ qubit and $\sigma_{\pm}=\frac{1}{2}(\sigma_{x}\pm i\sigma_{y})$. The whole expression of Hamiltonian represented by the matrices are given in appendix. 

The states is driven by the time-evolution operator $S(t,t_{0})$
\begin{eqnarray}
\label{stateevolution}
|\Psi(t)\rangle = S(t,t_{0}) |\Psi(t_{0})\rangle,
\end{eqnarray}
where $|\Psi(t)\rangle$ is the fermion occupation number states. The initial state $|\Psi(t_0)\rangle$ can be either one-particle state or multi-particle state which consists of several non-interacting particles and is prepared as a direct product of one-particle states. Breaking up the time interval from $t_0$ to $t$ into $n_{t}$ pieces of small duration $\Delta t$, the $S(t,t_{0})$ is represented in the Schr$\ddot{o}$dinger picture as
\begin{equation}
S(t,t_{0})=e^{-iH(t-t_0)}={\underbrace { S(t,t-\Delta t)\cdots  S(t_{0}+\Delta t,t_{0}) }_{n_{t}}},
\end{equation}
where $n_{t}\cdot\Delta t=t-t_{0}$.
The Trotter expansion of factor $S(t+\Delta t,t)\approx e^{-iH\Delta t}$ can be used to calculate the $S(t,t_{0})$. Finally, eq. (\ref{stateevolution}) can be approximated by series of quantum circuits that we give an example of them in FIG.~\ref{fig2}. 

We are interested in the state evolution depicted by the probability density $\rho_{s,i}(x,t)$
\begin{eqnarray}
\rho_{s,i}(x,t)&=&|\langle 1|_{x,s,i}\Psi(x,t)\rangle|^2=\langle \Psi(x,t)|1\rangle _{x,s,i}  \langle 1|_{x,s,i} \Psi(x,t)\rangle\nonumber\\
&=&\langle \Psi(x,t)|\hat{n}_{x,s,i}|\Psi(x,t)\rangle,
\end{eqnarray}
where $|1\rangle_{x,s,i}$ is defined in eq. (\ref{bdagger0}) and $|\Psi(t)\rangle$ is the evoluting state in (\ref{stateevolution}). In the qubit representation, the particle number operator expressed in (\ref{c3}) can be written as $\hat{n}_{x,s,i}=|1\rangle_{x,s,i}  \langle 1 |_{x,s,i}$.

We have worked out the quantum algorithm to simulate the real time dynamics of Gross-Neveu model non-perturbatively. As can be easily counted out from FIG.~\ref{fig2}, the number of CNOT operations with one site and one $\Delta t$ step is $186$. Then the time and space complexity of the algorithm running on quantum computers are $186\times n_{x}\times n_{t}\times n_{r}$ and $4n_{x}+1$, where $n_{x}$, $n_{t}$ and $n_{r}$ are numbers of space steps, time steps and the repeated running on quantum computers respectively. 

\section{Real time dynamics Simulation by MPS}
\label{result}
In what follows we will implement the simulation on a classical computer by applying the MPS routine. The default parameters are taken to be
\begin{eqnarray*}
r=1, \quad  \kappa=0.5,\quad \Delta t=0.1, \quad n_{x}=11,\quad n_{t}=150,  \\
 x=\kappa \cdot k, (k=0,\cdots,n_{x}), \quad t=l\cdot\Delta t, (l=0,\cdots n_{t}).
\end{eqnarray*}
The total number of qubits in the simulation is 45 (1 auxiliary qubit and $4\times n_{x}=44$ working qubits). The total number of CNOT operations and time complexity are $186\times n_{x}\times n_{t}=306900$, while the space complexity is $2^{(4\times n_{x}+1)}=2^{45}$. The Trotter expansion brings about $(\Delta t)^{2} \sim 0.01$ error and the MPS method has less than $10^{-9}$ error.

\begin{figure}
\begin{center}
\includegraphics[width= 83 mm]{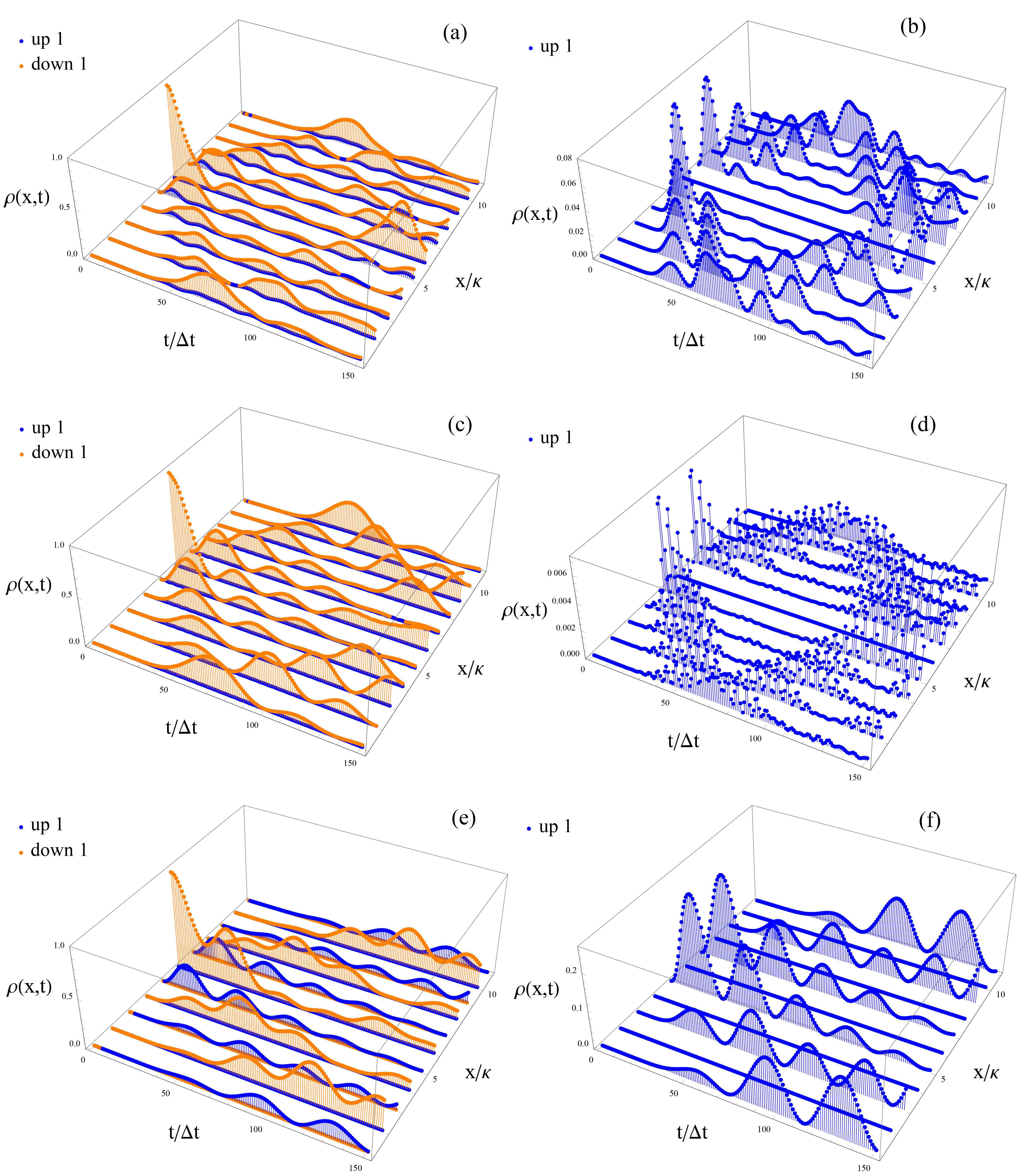}
\caption{The state evolutions on 11 sites in the duration of 150 time slices from an initial flavor-1 spin-down fermion sitting on site 6. The figures listed in the left column are the overall evolutions of the probability densities of the flavor-1 spin-up, flavor-1 spin-down, flavor-2 spin-up and flavor-2 spin-down (captioned as up 1, down 1, up 2 and down 2 in the figures) fermions with different parameters: (a) m=1, g=1.01, r=1; (c) m=10, g=1.01, r=1; (e) m=1, g=1.01, r=0. The probability densities of the two flavor-2 fermions are always zero in the evolution and have not been showned in the figures. The figures in the right column are the evolutions of flavor-1 spin-up fermion with the same parameters as their counterparts in the left column. }
\label{pic1}
\end{center}
\end{figure}

\begin{figure}[H]%{\color{white} H}
\begin{center}
\includegraphics[width= 83 mm]{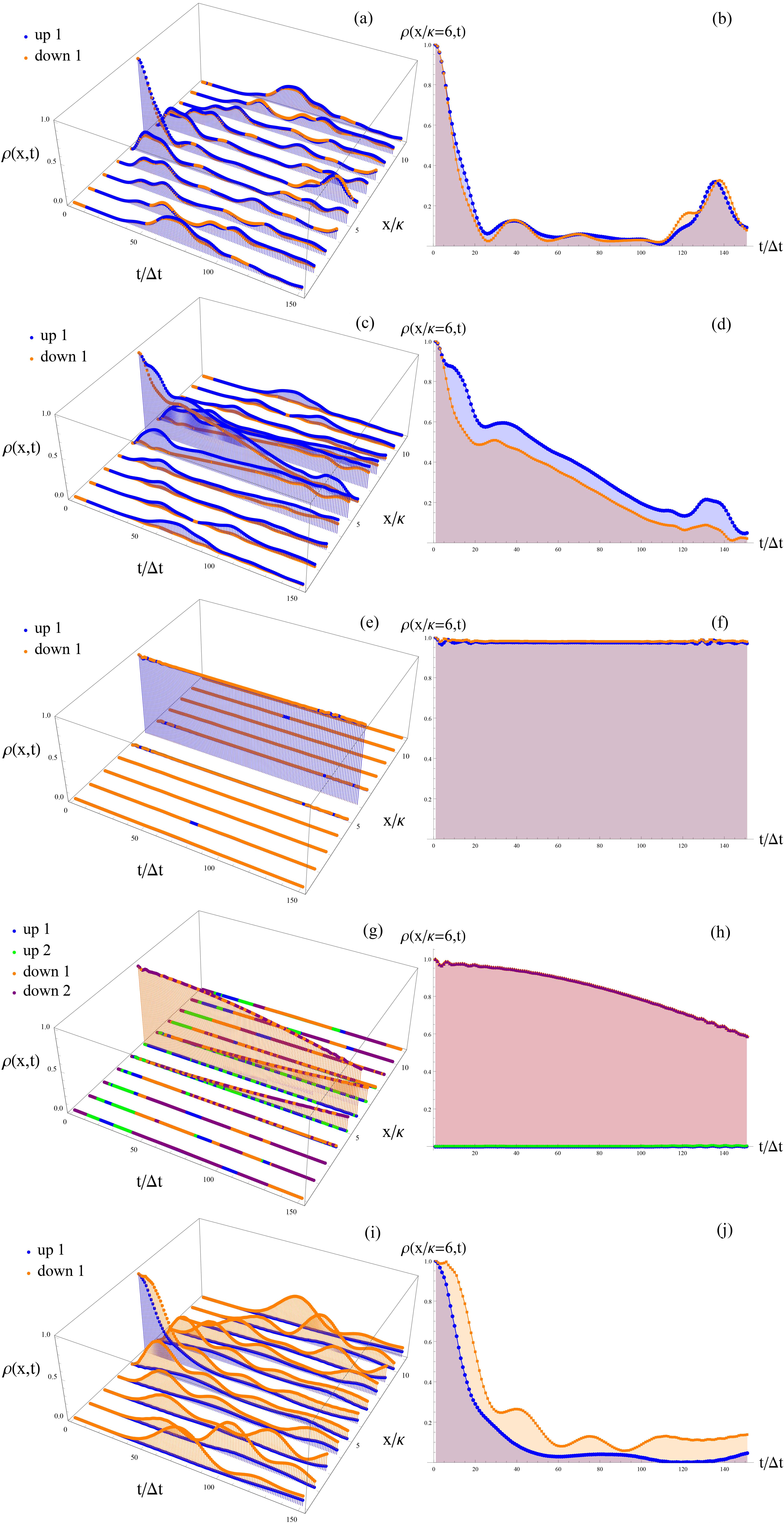}
\caption{The state evolutions on 11 sites in the duration of 150 time slices from an initial two-particle direct product state. The initial states of the Fig. (a), (c), (e) and (i) are composed by one flavor-1 spin-down and one flavor-1 spin-up fermions both located in site 6, while that of Fig. (g) is one flavor-1 spin-down and one flavor-2 spin-down fermion. The parameters of the model are (a) m=1, g=1.01; (c) m=1, g=3; (e) m=1, g=10; (g) m=1, g=10; (i) m=10, g=10. The figures in the right column are the evolutions of the site-6 particles in the corresponding figures of the left column.}
\label{pic2}
\end{center}
\end{figure}

\begin{figure}[H]
\begin{center}
\includegraphics[width= 83 mm]{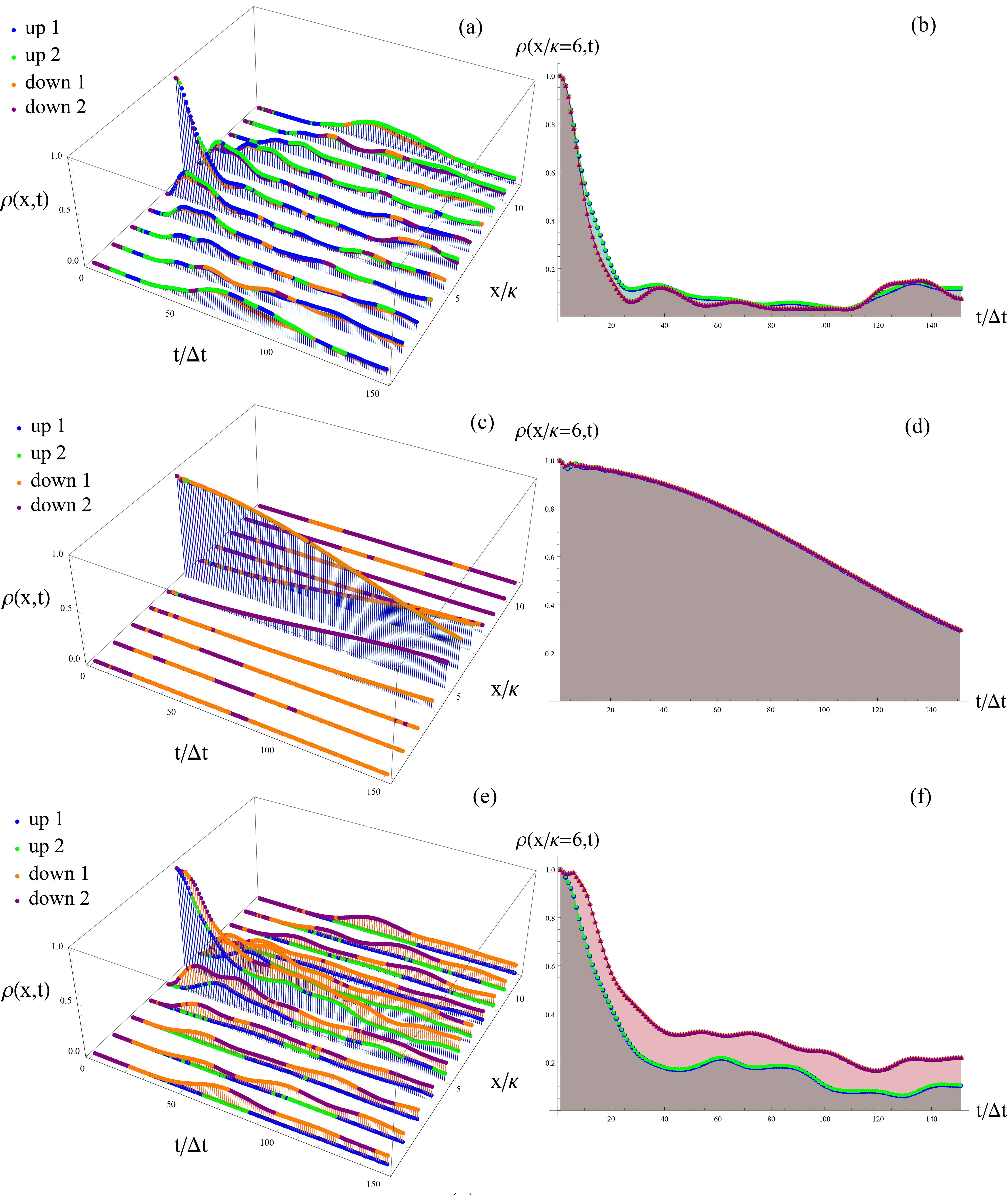}
\caption{The state evolutions on 11 sites in the duration of 150 time slices from an initial four-particle direct product state composed by flavor-1 spin-up, flavor-1 spin-down, flavor-2 spin-up and flavor-2 spin-down fermions located in site 6. The parameters of the model are (a) m=1, g=1.01; (c) m=1, g=10; (e) m=10, g=10. The figures in the right column are the evolutions of the site-6 particles in the left column.}
\label{pic3}
\end{center}
\end{figure}

Our major calculation results are presented in FIG.~\ref{pic1}, \ref{pic2} and \ref{pic3} which illustrate the probability density $\rho_{s,i}(x,t)$ of particle states with each flavor and spin as function of space and time. All of these figures depict the overall evolution of the four fermions on each site in the period of 150 time slices. These flavor-1 spin-up, flavor-1 spin-down, flavor-2 spin-up and flavor-2 spin-down fermions are captioned as up 1, down 1, up 2 and down 2 respectively in the figures. When a certain particle never shows up in the state evolution we do not include it in figures for a proper display, but it should keep in mind that the probability density of the particle is zero at every site and time. One can observe from the figures that the probability densities of particles propergate in space and time just like oscillation spreading out by wave in the water. We have applied a boundary condition that a wave is completely reflected when it propergates to site 0 and site 11. Thus each figure can be devided to two sections in time. One is from $t=0$ to about $t=50$, depiciting the state evolution from an initial state. The other section is a superposition of evolution from the initial state and the reflected states.

We begin with a simplest case of simulation and use it to depict how to read the figures properly. The initial state in FIG.~\ref{pic1} is a single flavor-1 spin-down particle located in site 6. The figures listed in the left column are the overall evolutions of the probability densities of the flavor-1 spin-up, flavor-1 spin-down, flavor-2 spin-up and flavor-2 spin-down fermions with different parameters: (a) m=1, g=1.01; (c) m=10, g=1.01; (e) m=1, g=1.01, r=0. The probability densities of the two flavor-2 fermions are always zero in the evolution and have not been showned in the figures. The flavor-1 spin-up fermion can be created and the figures in the right column are its evolutions with the same parameters as their counterparts in the left column. By comparing Fig.(a) and (b) with Fig.(c) and (d), one can find that the flavor-1 spin-up fermion is easier to be created when its mass is smaller. Fig.(e) and (f) is the case with fermion doubling problem which is not natural for fermion propagation with a step-size 2. 

We can find something interesting in FIG.~\ref{pic2}, where the state evolutions on 11 sites in the duration of 150 time slices from an initial two-particle direct product state. The figures in the left column are the evolutions on 11 sites and that in the right column are the evolutions of the site-6 particles in the counter left figures. The initial states of the Fig.(a), (c), (e) and (i) are composed by one spin-down and one spin-up fermions of flavor-1 both located in site 6. Fig.(a), (c) and (e) demonstrate the evolutions differ in coupling constant with $g=1.01, 3$ and $10$ respectively and all with the same fermion mass $m=1$. From these figures one can learn how the coupling constant makes effect to the state evolutions. With its value increases, the propagation of the initial particles diminishes, which can be observed more clear from the corresponding figures in the right column. Especially when in the case of strong coulping $g=10$, the interaction terms in eq.(\ref{hamiltonian}) is dominant. The particle states can be taken as near-eigenstates of the Hamiltonian. The two fermions almost keep stationary at the initial position as shown in Fig.(e) and (f). Comparing Fig.(e) with Fig.(i), both of which illustrate the evolution from the same initial state at the strong coupling $g=10$, one can find the effects of fermion mass. With a rather large fermion mass $m=10$, the mass term has an unignorable counterbalance to the coupling term in eq.(\ref{hamiltonian}). A visible portion of probability density can be found in Fig.(i) to transit from flavor-1 spin up fermion to flavor-1 spin down fermion and the two single fermions do not evolve in the same way anymore. If the bound states can be identified as fundamental particles confining in a lattice and evolving synchronistically, the two-particle state in Fig.(e) is a bound state while that in Fig.(i) is not. Then these observations are in agreement with the well known statement that bound states will appear in the particle spectrum when the mass of fundamental fermion is small enough and the coupling is strong enough \cite{Nambu1961tp, Nambu1961fr, gross1974dynamical}. In addition we also present the evolution from one flavor-1 spin-down and one flavor-2 spin-down fermion with the parameters $m=1$ and $g=10$ in Fig.(g) and (h), which show that the two fundamental fermions with the same spin of different flavor evolve in the same manner. This is a manifestation of the flavor symmetry.

The FIG.~\ref{pic3} are 150-time-slice evolutions out of an initial four particle direct product state composed by flavor-1 spin-up, flavor-1 spin-down, flavor-2 spin-up and flavor-2 spin-down fermions located in site 6. The parameters of the model in the figures are (a) m=1, g=1.01; (c) m=1, g=10; (e) m=10, g=10. One can see from all of the figures that the fundamental fermions with the same spin of different flavor evolve in the same manner, as is infered from the flavor symmetry. Especially in the Fig. (c) and (d) where the parameters are $m=1$ and $g=10$, the four fermions evolve almost in the same way and look like to form a bound state which propagates in space time.

\section{Summary and Conclusion}
\label{con}
We have presented a quantum algorithm to non-perturbatively simulate the two-flavor massive Gross-Neveu model in Schr$\ddot{o}$dinger picture. The simulation is implemented on a classic computer by applying the matrix product state representation. The real time evolutions of up to four particles on a site in initial state are figured out in space-time coordinate. The fermion mass and coupling play important roles in the state evolutions. Especially when the mass of fermion is small enough and the coupling is strong enough, the fundamental fermions evolve synchronistically in space from the two-fermion and four-fermion initial states. These are also the conditions on which the bound states made up of fundamental fermion pairs were found to arise automatically in the literatures. If what we observe in the two-fermion and four-fermion states are  bound states, this simulation might provide us an intuitive way to identify the bound states in view of fundalmental particles and images on bound state formation.  

\begin{acknowledgements}
This work is supported by the National Basic Research Program of China under Grant No. 2016YFA0301903 and the National Natural Science Foundation of China under Grant Nos. 11174370, 11304387, 61632021, 11305262, 61205108, and 11574398. We thank Zhi-Qin Zhang for picture drawing.
\end{acknowledgements}

%\clearpage
%\appendix
\section*{Appendix: The Pauli matrices formulations of Hamiltonians and the corresponding quantum circuits}
\begin{figure}[H]
\begin{center}
\includegraphics[width= 80 mm]{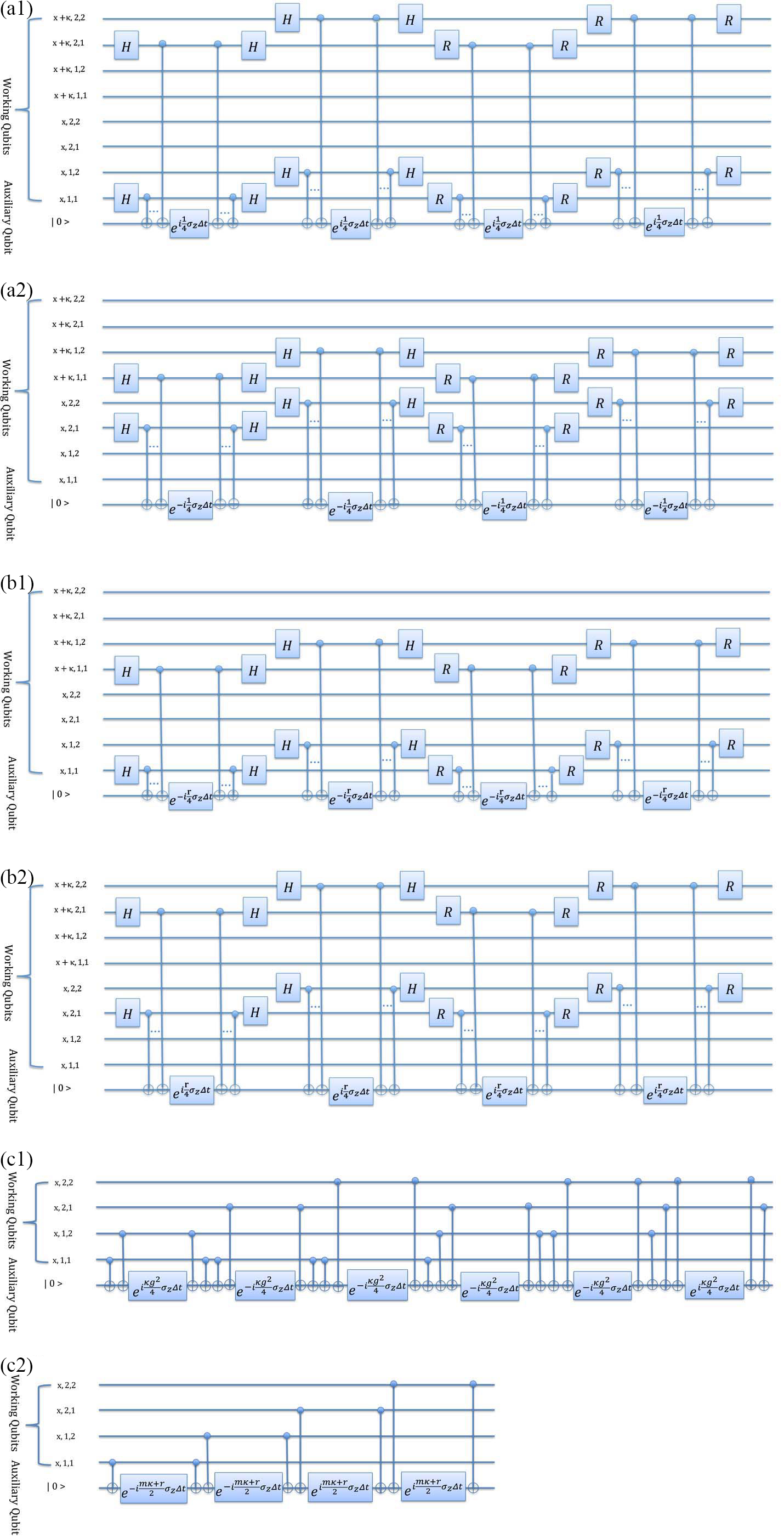}
\caption{Figs.~(a), (b) and (c) are quantum circuits to simulate unitary transformation $e^{-iH_{1}\Delta t}$, $e^{-iH_{2}\Delta t}$ and $e^{-iH_{3}\Delta t}$ respectively. The $H$ in the figures denotes Hadamard gate and $R$ is defined in \eqref{Rm}. }
\label{fig2}
\end{center}
\end{figure}

We devide the Hamiltonian into three terms
\begin{eqnarray}
H=\sum_{x\in \Omega,i=1,2 } \left(H_{1}+H_{2}+H_{3}\right).
\end{eqnarray}
The Pauli matric representation of the Hamiltonian is
\begin{eqnarray}  \nonumber
H_{1}&=&\frac{1}{2}(\hat{a}^{\dagger}_{x,i}\hat{b}_{x+\kappa,i}+\hat{b}^{\dagger}_{x+\kappa,i}\hat{a}_{x,i}-\hat{b}^{\dagger}_{x,i}\hat{a}_{x+\kappa,i}-\hat{a}^{\dagger}_{x+\kappa,i}\hat{b}_{x,i})\\  \nonumber
%&=&\frac{1}{2} \left(\sigma_{+}^{x,1,i}(\prod_{\alpha_{1}} -\sigma_{z}^{\alpha_{1}})\sigma_{-}^{x+\kappa,2,i}+\sigma_{+}^{x+\kappa,2,i}(\prod_{\alpha_{1}} -\sigma_{z}^{\alpha_{1}})\sigma_{-}^{x,1,i}\right.\\  \nonumber
%&&-\left.\sigma_{+}^{x,2,i}(\prod_{\bar{\alpha}_{1}} -\sigma_{z}^{\bar{\alpha}_{1}})\sigma_{-}^{x+\kappa,1,i}-\sigma_{+}^{x+\kappa,1,i}(\prod_{\bar{\alpha}_{1}} -\sigma_{z}^{\bar{\alpha}_{1}})\sigma_{-}^{x,2,i}
%\right)\\  \nonumber
&=&\frac{1}{4}\left[\sigma_{x}^{x,1,i}(\prod_{\alpha_{1}} -\sigma_{z}^{\alpha_{1}})\sigma_{x}^{x+\kappa,2,i}+\sigma_{y}^{x,1,i}(\prod_{\alpha_{1}} -\sigma_{z}^{\alpha_{1}})\sigma_{y}^{x+\kappa,2,i}\right.\\  \nonumber
&& \left.-\sigma_{x}^{x,2,i}(\prod_{\bar{\alpha}_{1}} -\sigma_{z}^{\bar{\alpha}_{1}})\sigma_{x}^{x+\kappa,1,i}-\sigma_{y}^{x,2,i}(\prod_{\bar{\alpha}_{1}} -\sigma_{z}^{\bar{\alpha}_{1}})\sigma_{y}^{x+\kappa,1,i}\right],
\end{eqnarray}

\begin{eqnarray}  \nonumber
H_{2}&=&\frac{r}{2}\left(-\hat{a}^{\dagger}_{x,i}  \hat{a}_{x+\kappa,i}-\hat{a}^{\dagger}_{x+\kappa,i}\hat{a}_{x,i} +\hat{b}^{\dagger}_{x,i}  \hat{b}_{x+\kappa,i}+\hat{b}^{\dagger}_{x+\kappa,i}\hat{b}_{x,i} \right)\\  \nonumber
%&=& \frac{r}{2}\left( -\sigma_{+}^{x,1,i}(\prod_{\alpha_{2}} -\sigma_{z}^{\alpha_{2}})\sigma_{-}^{x+\kappa,1,i} -\sigma_{+}^{x+\kappa,1,i}(\prod_{\alpha_{2}} -\sigma_{z}^{\alpha_{2}})\sigma_{-}^{x,1,i} \right.\\  \nonumber
%&&\left. +\sigma_{+}^{x,2,i}(\prod_{\bar{\alpha}_{2}} -\sigma_{z}^{\bar{\alpha}_{2}})\sigma_{-}^{x+\kappa,2,i} +\sigma_{+}^{x+\kappa,2,i}(\prod_{\bar{\alpha}_{2}} -\sigma_{z}^{\bar{\alpha}_{2}})\sigma_{-}^{x,2,i}\right)\\  \nonumber
&=&\frac{r}{4}\left[-\sigma_{x}^{x,1,i}(\prod_{\alpha_{2}} -\sigma_{z}^{\alpha_{2}})\sigma_{x}^{x+\kappa,1,i}-\sigma_{y}^{x,1,i}(\prod_{\alpha_{2}} -\sigma_{z}^{\alpha_{2}})\sigma_{y}^{x+\kappa,1,i}\right.\\  \nonumber
&& \left.+\sigma_{x}^{x,2,i}(\prod_{\bar{\alpha}_{2}} -\sigma_{z}^{\bar{\alpha}_{2}})\sigma_{x}^{x+\kappa,2,i}+\sigma_{y}^{x,2,i}(\prod_{\bar{\alpha}_{2}} -\sigma_{z}^{\bar{\alpha}_{2}})\sigma_{y}^{x+\kappa,2,i}\right],
\end{eqnarray}

\begin{eqnarray}\nonumber
H_{3}&=&(m\kappa+r)\left( \hat{n}_{x,up,i}- \hat{n}_{x,down,i}\right)-\frac{g^{2}\kappa}{2}(\hat{n}_{x,up,i}- \hat{n}_{x,down,i})^{2}\\  \nonumber
&=&\frac{m\kappa+r}{2} \left( \sigma_{z}^{x,1,1}+\sigma_{z}^{x,1,2}-\sigma_{z}^{x,2,1}-\sigma_{z}^{x,2,2}\right)\\  \nonumber
&&+\frac{ g^{2} \kappa}{4}\left( -2I-{\sigma_{z}^{x,1,1}} {\sigma_{z}^{x,1,2}}+{\sigma_{z}^{x,1,1}} {\sigma_{z}^{x,2,1}}+{\sigma_{z}^{x,1,1}} {\sigma_{z}^{x,2,2}}+\right.\\  \nonumber
&&\left.{\sigma_{z}^{x,1,2}} {\sigma_{z}^{x,2,1}}+{\sigma_{z}^{x,1,2}} {\sigma_{z}^{x,2,2}}-{\sigma_{z}^{x,2,1}} {\sigma_{z}^{x,2,2}}\right), 
\end{eqnarray}
where 
\begin{eqnarray*}
\alpha_{1}\in \{ (x^{\prime},s^{\prime},i^{\prime})|\mathcal{K}(x,1,i) \mathcal{K}(x^{\prime},s^{\prime},i^{\prime})<\mathcal{K}(x+\kappa,2,i)\},\\
\bar{\alpha}_{1}\in \{ (x^{\prime},s^{\prime},i^{\prime})|\mathcal{K}(x,2,i) \mathcal{K}(x^{\prime},s^{\prime},i^{\prime})<\mathcal{K}(x+\kappa,1,i)\},\\
\alpha_{2}\in \{ (x^{\prime},s^{\prime},i^{\prime})|\mathcal{K}(x,1,i) \mathcal{K}(x^{\prime},s^{\prime},i^{\prime})<\mathcal{K}(x+\kappa,1,i)\},\\
\bar{\alpha}_{2}\in \{ (x^{\prime},s^{\prime},i^{\prime})|\mathcal{K}(x,2,i) \mathcal{K}(x^{\prime},s^{\prime},i^{\prime})<\mathcal{K}(x+\kappa,2,i)\}.
\end{eqnarray*}
The quantum circuits to simulate the Hamiltonian evolution are plotted in FIG. \ref{fig2} where the $R$ gate is 
\begin{eqnarray}\label{Rm}
R=\frac{1}{\sqrt{2}}\left(\begin{array}{cc} 1& -i\\ i & -1 \end{array}\right)
\end{eqnarray}
with $R\sigma_{y}R=\sigma_{z}$.
\\

%\bibliography{mybibfile}   % name your BibTeX data base


%\bibliographystyle{apsrev4-1}
%
%\renewcommand{\baselinestretch}{1}
%\normalsize
%
%
%
%
%\clearpage%
%\phantomsection%
%\addcontentsline{toc}{chapter}{\numberline{}{Bibliography}}%
%\bibliography{mybibfile}


\end{document}